\documentclass[aps,nofootinbib,twocolumn,prd,eqsecnum,showpacs,showkeys,preprintnumbers]{revtex4-1}

\usepackage[caption=false]{subfig}
\usepackage{graphicx}
\usepackage{amsmath}
\usepackage{amsfonts}
\usepackage{amssymb}
\usepackage{color}
\usepackage{bm}
\usepackage{mathrsfs}
\usepackage{epstopdf}
\usepackage{url}
\usepackage{footnote}
\usepackage{textcomp}
\usepackage{dsfont}
\usepackage{ulem}

\makeatletter
\newcommand*{\rom}[1]{\expandafter\@slowromancap\romannumeral #1@}
\makeatother

\begin{document}

\title{Ghosts in the self-accelerating DGP branch with Gauss-Bonnet effect}
\author{Yen-Wei Liu $^{1,2}$}
\email{f97222009@ntu.edu.tw}
\author{Keisuke Izumi $^{1}$}
\email{izumi@phys.ntu.edu.tw}
\author{Mariam Bouhmadi-L\'{o}pez $^{3,4,5,6}$}
\email{{\mbox{mbl@ubi.pt. On leave of absence from UPV and IKERBASQUE.}}}
\author{Pisin Chen $^{1,2,7,8}$}
\email{chen@slac.stanford.edu}
\date{\today}

\affiliation{
${}^1$Leung Center for Cosmology and Particle Astrophysics, National Taiwan University, Taipei, Taiwan 10617\\
${}^2$Department of Physics, National Taiwan University, Taipei, Taiwan 10617\\
${}^3$Departamento de F\'{i}sica, Universidade da Beira Interior, 6200 Covilh\~a, Portugal\\
${}^4$Centro de Matem\'atica e Aplica\c{c}\~oes da Universidade da Beira Interior (CMA-UBI), 6200 Covilh\~a, Portugal\\
${}^5$Department of Theoretical Physics, University of the Basque Country UPV/EHU, P.O. Box 644, 48080 Bilbao, Spain\\
${}^6$IKERBASQUE, Basque Foundation for Science, 48011, Bilbao, Spain\\
${}^7$Graduate Institute of Astrophysics, National Taiwan University, Taipei, Taiwan 10617\\
${}^8$Kavli Institute for Particle Astrophysics and Cosmology, SLAC National Accelerator Laboratory, Stanford University, Stanford, California 94305, U.S.A.
}

\begin{abstract}
The Dvali-Gabadadze-Porrati brane-world model provides a possible approach to address the late-time cosmic acceleration. However, it has subsequently been pointed out that a ghost instability will arise on the self-accelerating branch. Here, we carefully investigate whether this ghost problem could be possibly cured by introducing the Gauss-Bonnet term in the five-dimensional bulk action, a natural generalization to the Dvali-Gabadadze-Porrati model. Our analysis is carried out for a background where a de Sitter brane is embedded in an anti--de Sitter bulk. Our result shows that the ghost excitations cannot be avoided even in this modified model.
\end{abstract}


\maketitle

\section{introduction} \label{introduction}

In recent years, the late-time cosmic acceleration has been confirmed by several observational evidences \cite{SNa,CMB,SDSS}. This important discovery leads to one of the great puzzles in cosmology, and various plausible models have been developed to unravel the nature of such a late-time speed-up over the last decade. There have been many attempts at building up reasonable and consistent models by modifying the standard cosmology, which can be roughly categorized into two major directions: one is to introduce a dominant dark energy component in the Universe (see, e.g., Ref.~\cite{Copeland:2006wr}), while the other is to modify Einstein's general relativity (GR) at large scales (see, e.g., Ref.~\cite{MG}).

An intriguing brane-world scenario proposed by Dvali, Gabadadze, and Porrati (DGP) provides a new mechanism with an induced gravity (IG) term, i.e., a four-dimensional (4D) Ricci scalar, included in the brane action \cite{Dvali:2000hr}. The IG term is expected to arise as a quantum correction due to the matter field on the brane \cite{Sakharov:1967pk}, and it makes possible to reproduce the correct 4D Newtonian gravity at short distances even if the bulk is a five-dimensional (5D) Minkowski space-time with an infinite size \cite{Dvali:2000hr}. The promising feature of the DGP model is that, when generalized to a Friedmann-Lema\^{\i}tre-Robertson-Walker (FLRW) brane with ordinary matter on it, one of its solutions, called the self-accelerating branch, will become asymptotically de Sitter in the far future, giving rise to a late-time accelerating phase without needing to introduce additional substances on the brane that violates the strong energy condition \cite{Shtanov:2000vr,Deffayet:2000uy}.

Despite this advantage, it was pointed out later on that the self-accelerating branch is plagued with a ghost instability \cite{Luty:2003vm,Nicolis:2004qq,Koyama:2005tx,Gorbunov:2005zk,Charmousis:2006pn,Koyama:2007za}. The spin-2 perturbations in this branch, viewed as an effective 4D massive gravity theory on a de Sitter background, are composed of a tower of infinite Kaluza-Klein (KK) massive gravitons. Then, the mass of the lowest mode $m$ is within the range $0<m^2<2H^2$ if the brane tension is positive, where $H$ is the Hubble parameter, and thus there will be a spin-2 ghost excitation in its helicity-0 component \cite{Higuchi:1986py}\footnote{
The instability of the ghost might be suppressed due to the spontaneous breaking of Lorentz symmetry by the helicity-0 ghost \cite{Izumi:2007pb,Izumi:2008st}.}. On the other hand, if the brane tension is negative, the lowest mass is larger than the critical scale, i.e., $2H^2<m^2$, but the spin-0 perturbation, associated with the brane-bending mode, becomes a ghost instead \cite{Koyama:2005tx}. In the specific case without brane tension, the lowest mass is equal to the critical scale. Even in this marginal case a detailed analysis shows the existence of a ghost from the mixing between the spin-0 sector and the helicity-0 part of the spin-2 sector \cite{Gorbunov:2005zk}. Furthermore, the appearance of ghosts in the DGP self-accelerating branch cannot be eliminated even by invoking a second brane in the bulk with a stabilization mechanism \cite{Izumi:2006ca}.  For more discussions on DGP ghosts, please see Ref.~\cite{Koyama:2007za} and the references therein. Nonlinear instabilities of the model have also been discussed in Refs.~\cite{Izumi:2007gs,Gregory:2007xy,Izumi:2014bba}.

In this paper, we will investigate the possibility of avoiding the ghost in a generalized DGP model. A natural generalization to the DGP gravitational action, based on its higher-dimensional nature, is by adding the Gauss-Bonnet (GB) term to the original 5D bulk action \cite{Lovelock:1971yv,Zwiebach:1985uq,Zumino:1985dp,Witten:1988hc,Chamseddine:1989nu,Zanelli:2005sa,deRham:2006pe,Kofinas:2003rz,
Brown:2006mh,BouhmadiLopez:2008nf,BouhmadiLopez:2009jk,BouhmadiLopez:2011xi,Belkacemi:2011zk,BouhmadiLopez:2012uf,Bouhmadi-Lopez:2013nma,
Bouhmadi-Lopez:2013gqa,Cho:2001su,Neupane:2001st,Neupane:2000wt,Neupane:2001kd,Germani:2012qm}. This modification then yields the most general field equation for the bulk metric with its derivatives only up to the second order \cite{Lovelock:1971yv}. Moreover, this GB term keeps the bulk theory ghost-free and arises as the leading-order correction to the low-energy effective action of the heterotic string theory \cite{Zwiebach:1985uq,Zumino:1985dp}, and furthermore, it plays an essential role in the Chern-Simons gauge theory of gravity \cite{Witten:1988hc,Chamseddine:1989nu,Zanelli:2005sa}.

This approach has been first discussed in Ref.~\cite{deRham:2006pe}.
Modes that have the potential to be ghosts are (i) the helicity-0 excitation of the lightest  KK graviton and (ii) the brane bending mode. In Ref.~\cite{deRham:2006pe}, considering a model with a minus sign in front of the bulk Einstein-Hilbert term, the authors succeeded in eliminating these modes.
However, considering only an Einstein-Hilbert term in the 5D action even with a minus sign generates ghost modes for the graviton in the bulk. The ghosts can be eliminated by introducing a GB term with a negative  prefactor and by taking the anti-de Sitter (AdS$_5$) vacuum called the GB vacuum. It is a kind of a ghost condensation of spin-2 particles.
Therefore, their model is ghost-free both in the five-dimensional theory and in the four dimensional effective theory.

The purpose of this paper is, in contrast, confirming if the straightforward extension of DGP model with the GB term still bothers the ghost mode.
As discussed in Ref.~\cite{deRham:2006pe},
we expect that the discrete light mode of the  KK tower of the graviton and the brane bending mode to appear in the four dimensional effective action.
Even though these two modes have the potential to be a ghost,
their existence does not necessarily imply a ghost excitation.
The massive graviton is a ghost excitation only if its mass squared is less than $2H^2$,
while the brane bending mode becomes a ghost only if the kinetic term has a wrong sign.
In the original DGP model, the ghost conditions have been carefully checked~\cite{Luty:2003vm,Nicolis:2004qq,Koyama:2005tx,Gorbunov:2005zk,Charmousis:2006pn,Koyama:2007za},
and one of the above mentioned condition is always satisfied, i.e. one of the ghost modes appears.
This means that the flip of the sign in the kinetic term of the brane bending mode happens when the squared value of the mass of the lightest KK graviton is $2H^2$. Nevertheless, its reason is still mysterious~\cite{Izumi:2006ca} and there is no reason why in the extended models the same happens.
Moreover, we may expect that the GB terms give a large correction to the self-accelerating branch.
With GB corrections, there are three branches of solutions; two branches appearing in the original DGP (the normal branch and the self-accelerating branch) and one additional branch called the GB branch.
The transition into the GB branch appears at the high energy region of the self-accelerating branch, and thus it is natural to expect that the self-accelerating branch is largely modified.
Therefore, a detailed analysis is needed to confirm the existence of the ghost.
Then, we study herein the linear perturbations around the background given by (\ref{metric}) and (\ref{warp}) with $\epsilon=+1$  which includes as a particular case the self-accelerating branch, and carefully examine whether or not the ghost excitations appearing in the DGP model could be possibly evaded in this framework. We show that even by including the GB term into the bulk, the ghost excitations are still present in this model.


The outline of the paper is as follows. In Sec.~\ref{model}, we consider the generalized DGP model with the GB term as well as a cosmological constant in the bulk. We then review the background solutions of this system. In Sec.~\ref{perturbation}, we study the linear perturbations over an AdS$_5$ bulk with a de Sitter brane within the model introduced in Sec.~\ref{model}. In Sec.~\ref{effaction}, we analyze the effective action for these perturbations, from which we examine the existence of the ghosts in this model. Finally, we present our summary in Sec.~\ref{summary}.

\section{The model}\label{model}

We consider a generalized DGP model with a GB term and a cosmological constant included in the bulk action. The gravitational action for this system is given by \cite{Kofinas:2003rz,Brown:2006mh,BouhmadiLopez:2008nf,BouhmadiLopez:2009jk,BouhmadiLopez:2011xi,Belkacemi:2011zk,BouhmadiLopez:2012uf,Bouhmadi-Lopez:2013nma,Bouhmadi-Lopez:2013gqa}
\begin{align}
S=&\frac1{2\kappa_5^2}\int_Md^5x\sqrt{-\,^{(5)\!}g}\left[\mathcal{R}-2\Lambda_5+\alpha\left(\mathcal{R}^2-4\mathcal{R}_{ab}
  \mathcal{R}^{ab}+\right.\right.\notag \\
  &\left.\left.\mathcal{R}_{abcd}\mathcal{R}^{abcd}\right)\right]-\frac1{\kappa_5^2}\int_{\Sigma_{\pm}}d^4x\sqrt{-g}\left[K+2\alpha\left(J-\right.\right.\notag\\
  &\left.\left.2G^{\mu\nu}K_{\mu\nu}\right)\right]+\int_{\Sigma}d^4x\sqrt{-g}\left[\frac{\gamma}{2\kappa^2_4}R-\lambda+\mathscr{L}_m\right]\label{action},
\end{align}
where the 5D manifold $M$ is split into two regions by a brane hypersurface $\Sigma$, and the two sides of the brane are denoted by $\Sigma_{\pm}$. The Latin indices $a,b,c,\ldots,$ run from 0 to 4, while the Greek indices $\mu,\nu,\ldots,$ run from 0 to 3. $^{(5)\!}g_{ab}$ is the five dimensional metric, and $g_{ab}=\,^{(5)\!}g_{ab}-n_an_b$ is the induced metric on the brane, with $n^a$ being the unit normal vector to the brane; $\mathcal{R}$, $R$, $\kappa_5^2$, $\Lambda_5\,(<0)$, $\lambda$, and $\mathscr{L}_m$ are the 5D Ricci scalar, the 4D Ricci scalar of the induced metric, the bulk gravitational constant, the bulk cosmological constant, the brane tension, and the matter Lagrangian on the brane, respectively. The GB parameter is denoted by $\alpha\,(\geq0)$, which has the dimension of length square, and the strength of the IG term is characterized by a dimensionless parameter $\gamma$. Moreover, the second term in Eq.(\ref{action}) corresponds to the generalized York-Gibbons-Hawking surface term \cite{York:1972sj,Gibbons:1976ue,Myers:1987yn,Davis:2002gn}, where $K_{\mu\nu}$ is the extrinsic curvature, $G_{\mu\nu}$ the Einstein tensor of the induced metric, and $J$ the trace of
\begin{align}
J_{\mu\nu}=&\frac13\left(2KK_{\mu\sigma}K^{\sigma}\,_{\nu}+K_{\rho\sigma}K^{\rho\sigma}K_{\mu\nu}\right.\notag\\
&\qquad
\left.-2K_{\mu\rho}K^{\rho\sigma}
K_{\sigma\nu}-K^2K_{\mu\nu}\right).
\end{align}

The 5D field equation, obtained by varying the bulk action in Eq.(\ref{action}), is given by \cite{Maeda:2003vq,Davis:2002gn,Dufaux:2004qs,Charmousis:2002rc,Maeda:2007cb}
\begin{equation}
\mathcal{G}_{ab}+\Lambda_5 \,^{(5)\!}g_{ab}-\frac{\alpha}2 \mathcal{H}_{ab}=0,\label{fieldeq}
\end{equation}
where $\mathcal{G}_{ab}$ is the 5D Einstein tensor and the quadratic curvature correction $\mathcal{H}_{ab}$ reads
\begin{align}
\mathcal{H}_{ab}=&\left(\mathcal{R}^2-4\mathcal{R}_{cd}\mathcal{R}^{cd}+\mathcal{R}_{cdef}
\mathcal{R}^{cdef}\right)\,^{(5)\!}g_{ab}\notag\\
&-4\left(\mathcal{R}\mathcal{R}_{ab}-2\mathcal{R}_{ac}\mathcal{R}_{b}\,^{c}-2\mathcal{R}_{acbd}\mathcal{R}^{cd}\right.\notag\\
&\left.+\mathcal{R}_{acde}\mathcal{R}_b\,^{cde}\right).
\end{align}
With the $\mathds{Z}_2$ symmetry assumed across the brane, the junction condition imposed at the brane is then given by \cite{Maeda:2003vq,Davis:2002gn,Dufaux:2004qs,Charmousis:2002rc,Maeda:2007cb},
\begin{align}
&K_{\mu\nu}+2\alpha\left[3J_{\mu\nu}-\frac23Jg_{\mu\nu}+\left(2P_{\mu\rho\sigma\nu}-\frac23g_{\mu\nu}G_{\rho\sigma}\right)K^{\rho\sigma}\right]\notag\\
&=r_c\bigg[-\kappa_4^2\left(T_{\mu\nu}-\frac13Tg_{\mu\nu}+\frac13\lambda g_{\mu\nu}\right)
+\gamma\left(G_{\mu\nu}-\frac13G g_{\mu\nu}\right)\bigg],\label{jc}
\end{align}
where the crossover scale $r_c$ is defined by $r_c\equiv\kappa_5^2/2\kappa_4^2$, $T_{\mu\nu}$ is the energy-momentum tensor of the matter content on the brane, and
\begin{align}
P_{\mu\nu\rho\sigma}=R_{\mu\nu\rho\sigma}+2R_{\nu[\rho}g_{\sigma]\mu}-2R_{\mu[\rho}g_{\sigma]\nu}+Rg_{\mu[\rho}g_{\sigma]\nu}\,\,.
\end{align}

Here, we consider a background solution corresponding to a de Sitter brane with a vanishing energy-momentum tensor of matter, i.e., $T_{\mu\nu}=0$, and the de Sitter brane is embedded in a bulk corresponding to two symmetric pieces of an AdS$_5$ space-time glued through the brane. In this configuration, the bulk field equation (\ref{fieldeq}) for the AdS$_5$ bulk implies the relation $\Lambda_5=-6\mu^2+12\alpha\mu^4$, where $\mu$ is the energy scale associated with the AdS$_5$ length $\ell\equiv1/\mu$ and has the following solutions:
\begin{equation}
\mu^2=\frac1{4\alpha}\left[1\pm\sqrt{1+\frac43\alpha\Lambda_5}\right].\label{mu}
\end{equation}
However, it has been proved that the bulk solution with the $+$ sign in Eq.(\ref{mu}) is perturbatively unstable \cite{Boulware:1985wk,Myers:1988ze,Cai:2001dz,Charmousis:2008ce}; therefore, we will focus on the $-$ branch from now on, and the energy scale $\mu^2$ is then bounded as\footnote{We have excluded the limiting case where $\mu^2=1/4\alpha$ corresponding to the Chern-Simons gravity, because in that case a homogeneous and isotropic brane cannot be embedded in a static bulk \cite{Charmousis:2002rc}. In addition, we assume a nonvanishing bulk cosmological constant $\Lambda_5$.} $0<\mu^2<1/4\alpha$ accordingly. Moreover, this background can be described by the bulk metric
\begin{equation}
ds^2=dy^2+n^2(y)\gamma_{\mu\nu}dx^\mu dx^\nu, \label{metric}
\end{equation}
where the brane is located at $y=0$, $\gamma_{\mu\nu}$ is the 4D de Sitter metric with its scalar curvature $R[\gamma_{\mu\nu}]=12H^2$, and the warp factor $n(y)$ is given by
\begin{equation}
n(y)=\frac H{\mu} \sinh{\left[\mu\left(y_*+\epsilon|y|\right)\right]},\label{warp}
\end{equation}
with
\begin{equation}
y_*=\frac1{\mu}\mathrm{arcsinh}{\left(\frac{\mu}H\right)},
\end{equation}
where the warp factor has two possible branches with $\epsilon=\pm1$ and is normalized at the position of the brane as $n(0)=1$.

\begin{figure}[tbp]
  \begin{center}
    \includegraphics[keepaspectratio=true,height=40mm]{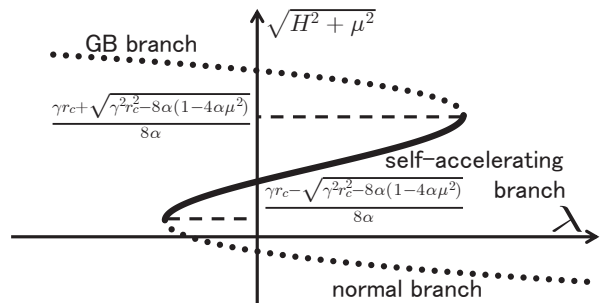}
  \end{center}
  \caption{   The relation between $\lambda$ and $\sqrt{H^2+\mu^2}$ for $\epsilon=+1$ in Eq.(\ref{Friedmann}):
  The solid line corresponds to the self-accelerating branch which is connected with the other two branches at one of the edges.
  The upper dotted line is called GB branch, while the lower branch corresponds to the normal branch with $\epsilon=+1$.}
  \label{fig:fig.pdf}
\end{figure}

From the junction condition (\ref{jc}), we derive the generalized Friedmann equation on the brane using the bulk metric (\ref{metric}),
\begin{align}
&\left[1+\frac83\alpha\left(H^2-\frac{\mu^2}2\right)\right]\sqrt{H^2+\mu^2}\notag\\
&=-\epsilon \,r_c\left(\frac{\kappa^2_4}{3}\lambda-\gamma H^2\right)\label{Friedmann}.
\end{align}
The relation between $\lambda$ and $\sqrt{H^2+\mu^2}$ is shown in fig.~\ref{fig:fig.pdf} and
the exact solutions of Eq.(\ref{Friedmann}) have been analyzed in Refs.~\cite{Kofinas:2003rz,Brown:2006mh,BouhmadiLopez:2008nf,BouhmadiLopez:2009jk,BouhmadiLopez:2011xi,Belkacemi:2011zk,
BouhmadiLopez:2012uf,Bouhmadi-Lopez:2013nma,Bouhmadi-Lopez:2013gqa}. In general,
there are three branches of solutions in Eq.(\ref{Friedmann}), among which the ``self-accelerating branch'' with $\epsilon=+1$ includes the DGP self-accelerating solution in the absence of the GB term and for a vanishing bulk cosmological constant \cite{Shtanov:2000vr,Deffayet:2000uy}, while the ``normal branch'' with $\epsilon=-1$, when switching off both the GB and IG effects, will recover the Randall-Sundrum single brane model.
We have an additional branch called the GB branch.
The transition to the GB branch appears at the high energy regime of the self-accelerating branch (see Fig.~\ref{fig:fig.pdf}).
The ghost problem is known to arise on the DGP self-accelerating branch, i.e., $\epsilon=+1$ with $\alpha\rightarrow0$ in Eq.(\ref{Friedmann}). We will therefore restrict our analysis to the
solution with $\epsilon=+1$ and $\alpha>0$, seeking to see if the GB term can in some way alleviate the ghost problem of the DGP self-accelerating model.

\section{Perturbed equations}\label{perturbation}

We now study the linear perturbations on the self-accelerating branch following the methodology used in Ref.~\cite{Izumi:2006ca} (see also Ref.~\cite{Yamashita:2014cra}). To derive the perturbed bulk field equation, we start with the perturbed metric
\begin{equation}
ds^2=dy^2+\left(n^2\gamma_{\mu\nu}+h^{TT}_{\mu\nu}\right)dx^\mu dx^\nu, \label{TT}
\end{equation}
with the transverse-traceless (TT) gauge-fixing conditions,
\begin{equation}
\nabla^{\mu}h^{TT}_{\mu\nu}=0, \quad h^{TT\mu}_{\,\,\,\mu}=0,
\end{equation}
where $\nabla_{\mu}$ is the covariant derivative associated with the background induced metric $g_{\mu\nu}=n^2\gamma_{\mu\nu}$. Using this TT perturbed metric (\ref{TT}), we obtain the perturbed bulk field equation from Eq.(\ref{fieldeq}),
\begin{align}
n^2h^{TT\prime\prime}_{\mu\nu}-2H^2h^{TT}_{\mu\nu}&-4n^2\mu^2h^{TT}_{\mu\nu}        \notag\\
&\,=-\left(\square-2H^2\right)h^{TT}_{\mu\nu},\label{pfieldeq}
\end{align}
where the d'Alembertian is defined by $\square=\gamma^{\mu\nu}\nabla_{\mu}\nabla_{\nu}$, and the prime denotes a partial derivative with respect to $y$. However, if we choose the TT gauge for the metric perturbations as in Eq.(\ref{TT}), the brane position cannot be fixed at $y=0$ but will in general reside at $y=\xi(x^{\mu})$ deviating slightly from the unperturbed position.

For the calculation of the perturbed junction condition, it is more convenient to introduce the Gaussian normal coordinate adapted to the brane hypersurface,
\begin{equation}
ds^2=d\bar y^2+\left(n^2\gamma_{\mu\nu}+\bar h_{\mu\nu}\right)d\bar x^\mu d\bar x^\nu, \label{GN}
\end{equation}
where the brane is now chosen to be fixed at $\bar y=0$. The perturbed junction condition imposed at the brane is then obtained by using this perturbed metric in Eq.(\ref{jc}),
\begin{align}
&\left(1-4\alpha\mu^2\right)\left(\bar{h}_{\mu\nu}^{\prime}-2\sqrt{H^2+\mu^2}\,\,\bar{h}_{\mu\nu}\right)=-\kappa^2_5\left(T_{\mu\nu}-\frac13\gamma_{\mu\nu}T\right)\notag\\
&\;\;+2\left(\gamma r_c-4\alpha\sqrt{H^2+\mu^2}\right)\left(X_{\mu\nu}-\frac13\gamma_{\mu\nu}X\right),\label{pjc}
\end{align}
where the energy-momentum tensor $T_{\mu\nu}$ is assumed to be first order, and
\begin{align}
X_{\mu\nu}&=-\frac12\left(\square\bar{h}_{\mu\nu}-\nabla_\mu\nabla_\rho\bar{h}^\rho_{\,\nu}-\nabla_\nu\nabla_\rho\bar{h}^\rho_{\,\mu}
+\nabla_\mu\nabla_\nu\bar{h}\right)\notag\\
&\!\!\!\!\!\!+\frac12\gamma_{\mu\nu}\left(\square\bar{h}-\nabla_\rho\nabla_\sigma\bar{h}^{\rho\sigma}\right)+H^2\left(\bar{h}_{\mu\nu}+\frac12\gamma_{\mu\nu}\bar{h}\right).
\end{align}

Now we have two perturbed equations expressed under two different gauge-fixing conditions. The gauge transformation between the metric written in the TT coordinate (\ref{TT}) and that in the Gaussian normal coordinate (\ref{GN}) can be carried out by the following coordinate transformations:
\begin{align}
y-\bar{y}&=\hat{\xi}^y(x^\rho),\label{xiy}\\
x^\mu-\bar{x}^\mu&=\frac{\sqrt{H^2+n^2\mu^2}}{nH^2}\gamma^{\mu\sigma}\partial_\sigma\hat{\xi}^y(x^\rho)+\hat{\xi}^\mu(x^\rho).
\label{xix}
\end{align}
Therefore, the transformation of the metric perturbations between these two coordinates is obtained as follows:
\begin{align}
\bar{h}_{\mu\nu}=&h^{TT}_{\mu\nu}+\frac{2\,n\sqrt{H^2+n^2\mu^2}}{H^2}\,\nabla_\mu\nabla_\nu\,\hat{\xi}^y\notag\\
&+2\,n\sqrt{H^2+n^2\mu^2}\,\gamma_{\mu\nu}\hat{\xi}^y+2\nabla_{(\mu}\hat{\xi}_{\nu)}.
\label{gauge1}
\end{align}
Furthermore, we can fix the function $\hat\xi^{\mu}$ such that the combination of the second and the last term in Eq.(\ref{gauge1}) vanishes at the position of the brane. As a result, after fixing the function $\hat\xi^{\mu}$ in this way, we have
\begin{align}
\bar{h}_{\mu\nu}=&h^{TT}_{\mu\nu}+\frac{2\,n}{H^2}\left(\sqrt{H^2+n^2\mu^2}-n\sqrt{H^2+\mu^2}\right)\nabla_\mu\nabla_\nu\,\hat{\xi}^y\notag\\
&+2\,n\sqrt{H^2+n^2\mu^2}\,\,\gamma_{\mu\nu}\hat{\xi}^y.
\label{gauge2}
\end{align}

Substituting the transformation (\ref{gauge2}) into the perturbed junction condition (\ref{pjc}), we can decompose the perturbed junction condition imposed at $y=0$ into the traceless part,
\begin{align}
&\left(1-4\alpha\mu^2\right)\left(h^{TT\prime}_{\mu\nu}-2\sqrt{H^2+\mu^2}\,h^{TT}_{\mu\nu}\right)+\left(\gamma r_c\vphantom{\sqrt{H^2+\mu^2}}\right.\notag\\
&\left.-4\alpha\sqrt{H^2+\mu^2}\right)\left(\square-2H^2\right)h^{TT}_{\mu\nu}=-\kappa^2_5\,\Sigma_{\mu\nu},\label{traceless}
\end{align}
where
\begin{align}
\Sigma_{\mu\nu}=&\left(T_{\mu\nu}-\frac14\gamma_{\mu\nu} T\right)-\frac2{\kappa^2_5}\left[1+4\alpha\left(\mu^2+2H^2\right)\vphantom{\sqrt{H^2+\mu^2}}\right.\notag\\
&\left.-2\gamma r_c\sqrt{H^2+\mu^2}\right]\left(\nabla_\mu\nabla_\nu-\frac14\gamma_{\mu\nu}\square\right)\hat{\xi}^y,
\label{source}
\end{align}
as well as the trace part,
\begin{eqnarray}
&&\left(\square+4H^2\right)\hat{\xi}^y \nonumber\\
&&\qquad
=\frac{\kappa^2_5}{6\left[2\gamma r_c\sqrt{H^2+\mu^2}-1-4\alpha\left(\mu^2+2H^2\right)\right]}T.
\nonumber\\
&& \label{trace}
\end{eqnarray}
$\hat \xi^y$ shows the position of the brane under TT gauge-fixing condition, and thus this equation represents the dynamics of the brane-bending mode. Moreover, it can be shown that the traceless source $\Sigma_{\mu\nu}$ also satisfies the transverse condition, $\nabla^{\mu}\Sigma_{\mu\nu}=0$, by using Eq.(\ref{trace}) and the following identity holding for any 4D scalar $F$:
\begin{equation}
\nabla^{\mu}\left(\nabla_{\mu}\nabla_{\nu}-\frac14\gamma_{\mu\nu}\square\right) F=\frac34n^{-2}\nabla_{\nu}
\left(\square+4H^2\right) F.
\end{equation}

We now proceed to derive the equations for the KK spin-2 modes and the brane-bending mode in the viewpoint of the 4D effective theory. The complete field equation for $h^{TT}$ can be obtained by combining its bulk part (\ref{pfieldeq}) and the junction condition imposed on it (\ref{traceless}),
\begin{align}
&\left(\hat{L}+\frac{\square-2H^2}{n^2}\right)h^{TT}_{\mu\nu}=-\frac2{1-4\alpha\mu^2}\left[\kappa^2_5\,\Sigma_{\mu\nu}+\left(\gamma r_c\vphantom{\sqrt{H^2+\mu^2}}\right.\right.      \notag\\
&\quad\left.\left.-4\alpha\sqrt{H^2+\mu^2}\right)\left(\square-2H^2\right)h^{TT}_{\mu\nu}\right]\delta(y),\label{hTT}
\end{align}
where the operator $\hat L$ is defined as
\begin{equation}
\hat L\equiv\frac1{n^2}\partial_yn^4\partial_y\frac1{n^2}.
\end{equation}
Eq.(\ref{hTT}) can be further separated into an eigenvalue problem through the KK decomposition of the metric perturbations $h^{TT}$,
\begin{equation}
h^{TT}_{\mu\nu}=\int dm\,h_{\mu\nu}^m(x^{\mu})\,u_m(y),\label{KK}
\end{equation}
where ``$\int$\,'' denotes a summation over the discrete modes and an integration over the continuous modes, and $u_m$ is the eigenfunction of the eigenvalue equation
\begin{align}
-\hat{L}\,u_m=&\frac{m^2}{n^2}\left[1+\frac2{1-4\alpha\mu^2}\left(\gamma r_c-4\alpha\vphantom{\sqrt{H^2+\mu^2}} \right.\right.\notag\\
&\left.\left.\times\sqrt{H^2+\mu^2}\right)\delta\left(y\right)\right]u_m,  \label{eigen}
\end{align}
where $m$ is the mass eigenvalue. Then, the field equation (\ref{hTT}), in terms of the KK decomposition (\ref{KK}), reduces to the simpler form,
\begin{align}
\int &dm\,\frac1{n^2}\left\{\left[1+\frac2{1-4\alpha\mu^2}\left(\gamma r_c-4\alpha\sqrt{H^2+\mu^2}\right)\delta(y)\right]\right.\notag\\
&\left.\vphantom{\frac2{1-4\alpha\mu^2}}\times\left(\square-2H^2-m^2\right)h^m_{\mu\nu}\,u_m\right\}=-\frac{2\,\kappa^2_5}{1-4\alpha\mu^2}\Sigma_{\mu\nu}\delta(y).\label{hm}
\end{align}
In addition, it can be shown that the eigenmodes $u_m$ in Eq.(\ref{eigen}) are mutually orthogonal with respect to the scalar product
\begin{align}
\left(u_{\tilde{m}},u_m\right)\equiv&\int^{\infty}_{-\infty}\frac{dy}{n^2}\bigg\{\left[1 -4\alpha\mu^2+2\left(\gamma r_c \vphantom{\sqrt{H^2+\mu^2}}\right.\right.\notag\\
&\left.\left.-4\alpha\sqrt{H^2+\mu^2}\right)\delta(y)\right]u_{\tilde{m}}u_m\bigg\}. \label{inner}
\end{align}
This definition of the scalar product (\ref{inner}) will always give rise to a positive number in the self-accelerating branch [see Eqs.(\ref{mu}) and (\ref{Friedmann})]. Therefore, we will use it here to normalize each eigenmode; i.e., the eigenmodes $u_m$ satisfy the condition $\left(u_{\tilde{m}},u_m\right)=\delta\left(\tilde{m},m\right)$, where the delta function $\delta\left(\tilde{m},m\right)$ is a Kronecker delta for the discrete modes and a Dirac delta function for a continuous spectrum. Notice that the scalar product defined in Eq.(\ref{inner}) is the same as that given in Ref.~\cite{Bouhmadi-Lopez:2013gqa} for the self-accelerating branch, since the eigenmodes defined in Ref.~\cite{Bouhmadi-Lopez:2013gqa}, $\mathcal{E}_{m}$, correspond to those defined here through $n^2\mathcal{E}_{m}=u_m$.

Given the scalar product constructed in Eq.(\ref{inner}), we can further simplify Eq.(\ref{hm}) by operating $\int^{\infty}_{-\infty}dy \,\tilde u_m$ on both sides of Eq.(\ref{hm}) and using the orthonormality of the eigenmodes. The resulting equation is then given by
\begin{equation}
\left(\square-2H^2-m^2\right)h^m_{\mu\nu}=-2\kappa^2_5\,\Sigma_{\mu\nu}\,u_m(0).\label{hm2}
\end{equation}
From Eqs.(\ref{KK}) and (\ref{hm2}), the solution for the metric perturbation $h^{TT}_{\mu\nu}$ can be written as
\begin{equation}
h^{TT}_{\mu\nu}=-2\kappa^2_5\int dm\,\frac{u_m(0)u_m(y)}{\square-2H^2-m^2}\Sigma_{\mu\nu}.\label{hTT2}
\end{equation}
Notice that $h^{TT}_{\mu\nu}$ is also sourced by the scalar mode $\hat\xi^y$ [see Eqs.(\ref{source}) and (\ref{trace})]. To obtain the full induced metric perturbations on the brane, we substitute Eqs.(\ref{trace}) and (\ref{hTT2}) back into Eq.(\ref{gauge2}) and evaluate the metric perturbation $\bar h_{\mu\nu}$ at $y=0$. In addition, we note that the gauge in the form of the Gaussian normal coordinate (\ref{GN}) is not yet completely fixed, so we can further fix this gauge freedom by eliminating the term proportional to $\nabla_{\mu}\nabla_{\nu}\,\hat\xi^{y}$. Therefore, after neglecting the term that can be erased by the gauge fixing, the induced metric perturbations on the brane, $\bar h^b_{\mu\nu}(x^\mu)\equiv\bar h_{\mu\nu}(x^\mu,0)$, is given by
\begin{widetext}
\begin{align}
\bar{h}^b_{\mu\nu}&(x^\mu)=-2\kappa_5^2\int dm \,u^2_m(0)\bigg\{\frac1{\square-2H^2-m^2}\left[\left(T_{\mu\nu}-\frac14\gamma_{\mu\nu}T\right)+\frac1{3(m^2-2H^2)}\left(\nabla_{\mu}
\nabla_{\nu}-\frac14\gamma_{\mu\nu}\square\right)T\right]\notag\\
&+\frac1{12(m^2-2H^2)}\gamma_{\mu\nu}T\bigg\}+\frac{\kappa^2_5}6\gamma_{\mu\nu}\left[\frac{2\sqrt{H^2+\mu^2}}{2\gamma r_c\sqrt{H^2+\mu^2}-1-4\alpha(\mu^2+2H^2)}+4H^2\left(\int dm\frac{u^2_m(0)}{(m^2-2H^2)}\right)\right]\frac1{\square+4H^2}T,\label{hb}
\end{align}
\end{widetext}
where we have already made use of the identity
\begin{eqnarray}
&&\frac1{\square-2H^2-m^2}\left(\nabla_{\mu}\nabla_{\nu}-\frac14\gamma_{\mu\nu}\,\square\right)F
\nonumber\\
&&\quad=\left(\nabla_{\mu}\nabla_{\nu}
\vphantom{\frac14}\right.
\,\,\left.-\frac14\gamma_{\mu\nu}\,\square\right)\frac1{\square+6H^2-m^2}\,F,
\end{eqnarray}
holding for an arbitrary 4D scalar $F$. Furthermore, we have also applied the operator decomposition as follows,
\begin{eqnarray}
&&\frac1{(\square+6H^2-m^2)(\square+4H^2)}\,F \nonumber\\
&&\quad
=\frac1{m^2-2H^2}\left(\frac1{\square+6H^2-m^2}-\frac1{\square+4H^2}\right)F,
\nonumber\\
&&\label{decomposition}
\end{eqnarray}
as long as $m^2\neq2H^2$. As can be seen from the resulting solution in Eq.(\ref{hb}), the physical degrees of freedom contained in the induced metric perturbations on the brane are effectively composed not only of a KK tower of massive spin-2 gravitons [cf.~Eq.(\ref{hFP2})] but of a spin-0 excitation associated with the brane-bending mode, corresponding to the second term of Eq.(\ref{hb}). In addition, we highlight that, if the trace of the energy-momentum tensor is nonzero, $T\neq0$, there is no pathology appearing in this model when $m^2=2H^2$, in contrast with the Fierz-Pauli model for the spin-2 field (see Appendix \ref{appendix}). In the Fierz-Pauli model, if the graviton mass square reaches the critical scale, $2H^2$, the constraint for the trace of the field, Eq.(\ref{constraintT}), implies that the trace of the energy-momentum tensor is zero. However, the seemingly divergent expression in Eq.(\ref{hb}) when $m^2=2H^2$ is simply because we have decomposed the result into the spin-2 and the spin-0 sector through Eq.(\ref{decomposition}), which does not hold if $m^2=2H^2$. In this sense, the spin-2 and the spin-0 sector are degenerate at this critical scale. Moreover, we can also check that there are no divergent expressions in Eqs.(\ref{trace}) and (\ref{hTT2}) when $m^2=2H^2$ before arriving at the final result (\ref{hb}).

\section{Effective action and existence of ghosts}\label{effaction}

To check explicitly the existence of ghosts in this model, we need to construct the 4D effective action responsible for all the degrees of freedom included in the solution (\ref{hb}). As we mentioned in the previous section, the induced metric perturbations (\ref{hb}) in general consist of a massive KK tower of spin-2 modes as well as a spin-0 mode, which are denoted here by $\bar h^{m}_{\mu\nu}$ and $s$, respectively. Therefore, the induced metric perturbations $\bar h^b_{\mu\nu}$ can be expressed as
\begin{equation}
\bar{h}^b_{\mu\nu}=\int dm\,\bar{h}^m_{\mu\nu}+\frac14\gamma_{\mu\nu}s.
\end{equation}
In terms of these notations, the lowest order action for the coupling of the matter trapped on the brane to the induced gravitons is then written as
\begin{eqnarray}
S_{\textrm{m}}&=&\frac12\int d^4x\sqrt{-\gamma}\,\bar{h}^{b}_{\mu\nu}\,T^{\mu\nu} \nonumber\\
&=&\frac12\int dm\int d^4x\sqrt{-\gamma}\,\bar{h}^{m}_{\mu\nu}\,T^{\mu\nu}
+\frac18\int dx^4\sqrt{-\gamma}\,s\,T. \nonumber\\
&&\label{Sm}
\end{eqnarray}

Now we continue to deduce the kinetic part of the effective action, and we will deal with the spin-2 and the spin-0 sector separately for convenience. For the spin-2 modes alone, we temporarily consider the traceless source here for simplicity, i.e., $T=0$, and so are the spin-2 fields $\bar h^m_{\mu\nu}$ (see Appendix \ref{appendix} for the general case without this restriction). Then, the kinetic part of the effective action for these TT spin-2 fields $\bar h^m_{\mu\nu}$ takes the form
\begin{equation}
S_{h}=\int dm\,\alpha_m\int d^4x\sqrt{-\gamma}\,\bar{h}^{m\mu\nu}(\square-2H^2-m^2)\bar{h}^{m}_{\mu\nu},\label{Sh}
\end{equation}
where the undetermined coefficients $\alpha_m$ can be fixed by comparing the equation of motion derived from the action $S_h+S_{\textrm{m}}$ with the one given in Eq.(\ref{hb}) as we consider only the spin-2 part with $T=0$. As a result, we have
\begin{equation}
\alpha_m=\frac1{8\kappa^2_5u_m^2(0)}.
\end{equation}
On the other hand, for the spin-0 mode alone, its kinetic part of the effective action is given by
\begin{equation}
S_{s}=\beta_s\int d^4x\sqrt{-\gamma}\,s(\square+4H^2)s,\label{Ss}
\end{equation}
where the coefficient $\beta_s$ can be fixed as well using the same method, i.e., matching the field equation derived from the action $S_s+S_{\textrm{m}}$ (only taking into account the second term of $S_{\textrm{m}}$ here) with only the spin-0 part of Eq.(\ref{hb}). Then, we end up with the result,
\begin{widetext}
\begin{equation}
\beta_s=-\frac3{64\kappa_5^2}\left[\frac{\sqrt{H^2+\mu^2}}{2\gamma r_c\sqrt{H^2+\mu^2}-1-4\alpha(\mu^2+2H^2)}+2H^2\left(\int dm\frac{u^2_m(0)}{(m^2-2H^2)}\right)\right]^{-1}.\label{betas}
\end{equation}
\end{widetext}

For the massive KK modes in this framework, it can be shown that the zero mode with $m=0$ satisfying the eigenvalue equation (\ref{eigen}) is not normalizable with respect to the scalar product (\ref{inner}). Therefore, as in the DGP model, there is no physically admissible zero mode in the self-accelerating branch here. However, it has been shown that, as mentioned in the Introduction, the helicity-0 component of the spin-2 field becomes a ghost if its mass is in the range $0<m^2<2H^2$ \cite{Higuchi:1986py}. Thus, if the lightest massive KK mode in this model has a mass in this forbidden range, this system will contain a spin-2 ghost. On the contrary, if the mass of the lightest KK mode is higher than the critical scale, i.e., $2H^2<m^2$, the spin-2 perturbations become healthy; however, whether the spin-0 mode is healthy in this case has yet to be checked.

We now turn to the spin-0 mode in the case where the mass of the lightest KK mode satisfies $2H^2<m^2$. The coefficient
of the action for the spin-0 mode, $\beta_s$, is explicitly given in Eq.(\ref{betas}), in which the second term inside the bracket is always positive in this case. Besides, it is known that,
from Fig.~\ref{fig:fig.pdf},
the self-accelerating branch of this model is restricted to the parameter space
\begin{align}
&\frac{\gamma r_c-\sqrt{\gamma^2r_c^2-8\alpha(1-4\alpha\mu^2)}}{8\alpha}<\sqrt{H^2+\mu^2} \notag\\
&<\frac{\gamma r_c+\sqrt{\gamma^2r_c^2-8\alpha(1-4\alpha\mu^2)}}{8\alpha},\label{range}
\end{align}
where the upper and lower bounds come from the transition to the GB branch and the normal branch but with $\epsilon=+1$, respectively.
Out of the range (\ref{range}) with $\epsilon=+1$, since the Hubble parameter is a decreasing function of the energy density, it cannot describe our Universe~\cite{BouhmadiLopez:2012uf,Bouhmadi-Lopez:2013nma,Bouhmadi-Lopez:2013gqa}.
Within the region of the above parameter space, the first term in the bracket of Eq.(\ref{betas}) turns out to be a positive value. As a result, the coefficient $\beta_s$ is always negative as long as the mass of the lowest mode is larger than the critical scale, i.e., $2H^2<m^2$, indicating the presence of a spin-0 ghost. Consequently, similarly to the DGP model, there always exists a ghost in the self-accelerating branch of this model, which implies that the ghost instability present in the DGP model still cannot be removed by invoking a GB term in the bulk action.

The surprising fact is that the sign flip of the first term in the bracket of Eq.(\ref{betas})
always happens with the transition of branches, that is, they are mysteriously related.
The lower bound of the inequality (\ref{range}) has been already found in the original DGP model, while the upper bound the  inequality (\ref{range}) is shown here for the first time. Then, although the GB branch (with $\epsilon=+1$) is a theoretical object and can never describe our Universe, it is interesting that this  branch can be ghost-free. Indeed, the GB branch despite having both modes (helicity-0 of the spin-2 sector and the brane bending mode) can be ghost free, unlike the self-accelerating branch where either the brane bending or the helicity-0 of the spin-2 is a ghost. The normal branch has a similar behaviour to that of the GB branch.
In summary, it  might be possible to construct a viable cosmological model where despite the existence of the brane bending and the helicity-0 modes, there is no ghost. This
 may imply the possibility of a ghost-free interesting solution for $\epsilon=+1$ with a nontrivial modification of the DGP model.

\section{summary}\label{summary}

In this paper, we looked into a generalized DGP brane-world scenario with a GB term as well as a cosmological constant both incorporated in the bulk action. To check whether this framework can possibly provide a way out of the DGP ghost instability, we have studied the linear perturbations around a de Sitter self-accelerating brane embedded in an AdS$_5$ bulk space-time. Having the linear perturbations analyzed in this system, we end up with the effective induced metric perturbations on the brane, Eq.(\ref{hb}), the physical degrees of freedom in which, as long as none of the KK modes has a critical mass, $m^2\neq2H^2$, can be effectively described in terms of the massive KK tower of the spin-2 gravitons as well as the spin-0 excitation associated with the brane-bending mode. Moreover, in contrast with the Fierz-Pauli model for the spin-2 field, gravity in this system can couple to matter with nonzero trace of the energy-momentum tensor when $m^2=2H^2$, at which the spin-2 and the spin-0 perturbations are degenerate. Therefore, one can no longer divide the degrees of freedom into the spin-2 and the spin-0 sector at this critical scale.

It has been shown that the massive spin-2 field contains a ghost excitation in its helicity-0 component if the mass is in the range $0<m^2<2H^2$ \cite{Higuchi:1986py}. Thus, from the massive gravity theory viewpoint, if the mass of the lightest KK mode here is within this forbidden range, there will be a spin-2 ghost excitation present in this system. On the other hand, provided that the mass of the lightest mode becomes larger than the critical scale, $2H^2<m^2$, although the spin-2 sector becomes healthy in this case, the spin-0 mode is shown to be a ghost instead, similarly to the DGP model. For the specific case where the lightest mass is equal to the critical scale, $m^2=2H^2$, whether or not a ghost exists in this model cannot be verified rigorously through the method we used here. Presumably, this model still contains a ghost in this marginal case as happens in the DGP model \cite{Gorbunov:2005zk}. However, this specific fine-tuning condition, i.e., $m^2=2H^2$, is easily broken provided that we consider physical matter fields on the brane, in which case the Hubble parameter in general varies with time. As a result, the DGP ghost instability at the level of linear perturbations still cannot be eliminated by invoking the GB term in the bulk action.

Our result shows that in the self-accelerating branch the ghost mode always appears, while in the other branches (with $\epsilon=+1$) we have the ghost-free parameter range, although these branches cannot describe our Universe.
This is because the sign of the kinetic term of the brane bending mode not only depends on the mass of the lightest KK graviton but also seems to be related to the branches.
We can see it from Eq.(\ref{betas}).
The sign of the first term in the bracket of Eq.(\ref{betas}) is positive for the self accelerating branch and negative for the other two branches.
Therefore, the brane bending mode has both informations of the branch and the value of the lightest KK mass.
This relation might be important for the future understanding of the origin of the ghost.
Finally, could other generalizations of the DGP model along the line of Refs.~\cite{BouhmadiLopez:2009db,BouhmadiLopez:2010pp} appease the ghost problem? We leave this question to a future work.

\acknowledgments
Y.W.L.~is supported by Taiwan National Science Council (TNSC) under Project No.~NSC 97-2112-M-002-026-MY3. K.I.~is supported by Taiwan National Science Council (TNSC) under Project No.~NSC101-2811-M-002-103. M.B.L.~is supported by the Basque Foundation for Science IKERBASQUE. She also wishes to acknowledge the hospitality of LeCosPA Center at the National Taiwan University during the completion of part of this work and the support of the Portuguese Agency ``Funda\c{c}\~{a}o para a Ci\^{e}ncia e Tecnologia" through PTDC/FIS/111032/2009. This work was partially supported by the Basque government Grant No.~IT592-13. P.C.~is supported by Taiwan National Science Council (TNSC) under Project No.~NSC 97-2112-M-002-026-MY3, by Taiwan National Center for Theoretical Sciences (NCTS), and by U.S. Department of Energy under Contract No. DE-AC03-76SF00515.


\appendix
\section{SPIN-2 FIELD ON THE DE SITTER BACKGROUND}\label{appendix}

The Lagrangian for the spin-2 field with the Fierz-Pauli mass term on the de Sitter background up to the second order is given by \cite{Higuchi:1986py}:
\begin{align}
L_{\textrm{FP}}&=\frac{1}{2\kappa_4^2}\sqrt{-g}\bigg[-\frac14\nabla_{\mu}h_{\nu\lambda}\nabla^{\mu}h^{\nu\lambda}+\frac12\nabla_{\mu}h^{\mu\lambda}
\nabla^{\nu}h_{\nu\lambda}\notag\\
&+\frac14\left(\nabla^{\mu}h-2\nabla_{\nu}h^{\mu\nu}\right)\nabla_{\mu}h-\frac{H^2}2\left(h_{\mu\nu}h^{\mu\nu}+\frac12h^2\right)\notag\\
&-\frac{m^2}4\left(h_{\mu\nu}h^{\mu\nu}-h^2\right)\bigg].\label{LFP}
\end{align}
The lowest-order Lagrangian for the coupling of gravity to matter on this background can be described by
\begin{equation}
L_\textrm{m}=\frac12\sqrt{-g}\,\,T^{\mu\nu} h_{\mu\nu}\label{Lm}.
\end{equation}
From Eqs.(\ref{LFP}) and (\ref{Lm}), the equation of motion for the spin-2 field, $h_{\mu\nu}$, is then obtained by varying the Lagrangian $L=L_{\textrm{FP}}+L_\textrm{m}$,
\begin{equation}
X_{\mu\nu}-\frac{m^2}2\left(h_{\mu\nu}-g_{\mu\nu}h\right)+\kappa_4^2 T_{\mu\nu}=0,\label{eomFP}
\end{equation}
where
\begin{align}
X_{\mu\nu}&=\frac12\left(\square h_{\mu\nu}-\nabla_{\mu}\nabla^{\sigma}h_{\sigma\nu}-\nabla_{\nu}\nabla^{\sigma}h_{\sigma\mu}+
\nabla_{\mu}\nabla_{\nu}h\right)\notag\\
&+\frac12g_{\mu\nu}\left(\nabla_{\alpha}\nabla_{\beta}h^{\alpha\beta}-\square h\right)-H^2\left(h_{\mu\nu}+\frac12g_{\mu\nu}h\right).
\end{align}
One can simply check that $X_{\mu\nu}$ satisfies the transverse condition, $\nabla^{\mu}X_{\mu\nu}=0$. With this transverse condition as well as the conservation condition, $\nabla^\mu T_{\mu\nu}=0$, we obtain the following constraints [see Eq.(\ref{eomFP})]:
\begin{equation}
\nabla^{\mu}h_{\mu\nu}=\nabla_{\nu}h.\label{constraints}
\end{equation}
Substituting these constraints (\ref{constraints}) back into the equation of motion (\ref{eomFP}), we have another constraint for the trace of the field:
\begin{equation}
h=-\frac{2\kappa^2_4}{3(m^2-2H^2)}T.\label{constraintT}
\end{equation}
Notice that if $m^2=2H^2$ here, the constraint (\ref{constraintT}) implies that $T=0$. As a result, having imposed all the constraints (\ref{constraints}) and (\ref{constraintT}) in the equation of motion (\ref{eomFP}), we finally have
\begin{align}
\left(\square-2H^2-m^2\right)h_{\mu\nu}=&-2\kappa_4^2T_{\mu\nu}+\left[\nabla_{\mu}\nabla_{\nu}\vphantom{\left(m^2-H^2\right)}\right.\notag\\
&\left.-\left(m^2-H^2\right)g_{\mu\nu}\right]h.\label{hFP1}
\end{align}
More clearly the field $h_{\mu\nu}$ in Eq.(\ref{hFP1}) can be further expressed in a form separated into a traceless and a trace part,
\begin{align}
h_{\mu\nu}=&-2\kappa^2_4 \bigg\{\frac1{\square-2H^2-m^2}\bigg[\bigg(T_{\mu\nu}-\frac14g_{\mu\nu}T\bigg)\notag\\
&+\frac1{3(m^2-2H^2)}\bigg(\nabla_{\mu}\nabla_{\nu}-\frac14g_{\mu\nu}\square\bigg)T\bigg]\notag\\
&+\frac1{12(m^2-2H^2)}g_{\mu\nu}T\bigg\}.\label{hFP2}
\end{align}

\end{document}